\documentclass[]{spie}
 
\usepackage[]{graphicx}
\usepackage{amsfonts}
\usepackage{mathtools}
\usepackage{mathptmx}
\usepackage{subfig}
\usepackage{floatrow}

\usepackage{sidecap}
\title{Registration of Longitudinal Spine CTs for Monitoring Lesion Growth}

\author{Malika Sanhinova\supit{a,b}, Nazim Haouchine\supit{a,c}, Steve D. Pieper\supit{d}, William M. Wells, III\supit{a,c}, Tracy A. Balboni\supit{a,c,e}, Alexander Spektor\supit{a,e}, Mai Anh Huynh\supit{a,e}, Jeffrey P. Guenette\supit{a,c}, Bryan Czajkowski\supit{a,c}, Sarah Caplan\supit{a,e}, Patrick Doyle\supit{a,e}, Heejoo Kang\supit{a,e}, David B. Hackney\supit{a,b}, Ron N. Alkalay\supit{a,b}
\skiplinehalf
\supit{a} Harvard Medical School, Boston, MA, USA; \\
\supit{b} Beth Israel Deaconess Medical Center, Boston, MA, USA \\
\supit{c} Brigham and Women's Hospital, Boston, MA, USA; \\
\supit{d} Isomics, Inc., Cambridge, MA, USA; \\
\supit{e} Dana-Farber Cancer Institute, Boston, MA, USA
}

\begin{document} 
\maketitle 

\begin{abstract}
Accurate and reliable registration of longitudinal spine images is essential for assessment of disease progression and surgical outcome. Implementing a fully automatic and robust registration is crucial for clinical use, however, it is challenging due to substantial change in shape and appearance due to lesions. In this paper we present a novel method to automatically align longitudinal spine CTs and accurately assess lesion progression. Our method follows a two-step pipeline where vertebrae are first automatically localized, labeled and 3D surfaces are generated using a deep learning model, then longitudinally aligned using a Gaussian mixture model surface registration. We tested our approach on 37 vertebrae, from 5 patients, with baseline CTs and 3, 6, and 12 months follow-ups leading to 111 registrations. Our experiment showed accurate registration with an average Hausdorff distance of 0.65 mm and average Dice score of 0.92. 
\end{abstract}


\section{DESCRIPTION OF PURPOSE}
\label{sec:intro}
Longitudinal lesion monitoring plays a pivotal role in observing changes of metastatic vertebrae and disease progression over time. By consistently evaluating the size, shape, and characteristics of lesions, healthcare professionals can gain valuable insights into the effectiveness of treatments, predict potential complications, and make informed decisions about patient care. The primary objective of this research is to conduct a longitudinal study focused on lesion growth in metastatic spines. Our approach allows for a comprehensive understanding of how metastatic lesions evolve and impact a patient's clinical course, ultimately leading to more personalized and effective medical interventions. The accurate tracking of lesions necessitates the alignment of 3D images captured at distinct time points. An initial CT scan establishes a baseline at 0 months, and follow-up scans at 3, 6, and 12-month are registered to this baseline. Registration is required due to altered parameters of the CT scan environment, such as origin, resolution, reconstruction methods, field of view, etc. This registration process enables lesion comparison and observation of the lesion development over time. 

Several notable contributions have been made in the field of advancing medical image registration techniques, each with its own limitations that provide valuable insights for further refinement and development. The methods described by Hille et al. \cite{Hille2018}, Cai et al. \cite{Cai2021} and Gueziri et al. \cite{Gueziri2019} address medical image registration involving MRI, CT, and ultrasound during spine interventions. These techniques are primarily utilized to fuse intraoperative images of lower resolution with pre-operative high-resolution scans, resulting in improved image quality—typically outperforming the outcomes of purely intraoperative CT imaging. However, while such registration methods exhibit speed and precision for surgical interventions, they presume a consistent appearance of vertebrae pre- and post-intervention, which might not align with cases involving lesions. Balakrishnan et al.\cite{Balakrishnan2018} and Hu et al. \cite{Hu2018} developed state-of-the-art learning-based registration strategies that integrate weakly-supervised segmentations during training, by incorporating them into the loss function. Zhao et al. \cite{Zhao2023} introduced SpineRegNet which allows for affine-elastic deformation field estimation for spine scans. By registering MRI to CT scans, the method combines rigid vertebral and elastic disk registration, ensuring the preservation of spine biomechanics. The framework integrates multiple modules for flexible spinal movement, fusion of multiple deformation fields, and preservation of vertebrae rigidity.

Previous methods are focused on pre-operative to intraoperative registration. To the best of our knowledge, longitudinal registration was first addressed by Glocker et al. in 2014 \cite{Glocker2014}. To overcome the initialization challenges of standard registration techniques for cases with small overlap, the authors proposed a registration method that includes a prior learning-based classification to estimate vertebrae centroids. This additional semantic information significantly improved registration compared to other initialization techniques, while applying the same intensity-based registration components. However, using intensity-based registration methods could be limiting for cases of metastatic spines. Metastatic vertebrae can undergo significant changes within a short time period, leading to dramatic alterations in their appearance and, consequently, intensity. This could hinder the accuracy and effectiveness of the intensity-based registration process. 

Our approach was developed to handle cases with large shape and intensity variations by utilizing segment-based registration. First, segmentation masks are predicted with a deep learning segmentation model that was trained on metastatic spines. Training set included spines with significant lesion progression and missing portions of vertebrae. Then, obtained segmentation masks are used to register follow-ups into a baseline on vertebra level. Rigid registration prevents deformation of bone structure, while segment-level registration ensures a unique transformation matrix per vertebra. Thus, the proposed two-step approach can be used for metastatic vertebrae undergoing significant shape and intensity changes. 

\section{METHODS}
\subsection{Multi-class Segmentation of Lesioned Vertebrea from CT}
Let us define the training set $T^\mathrm{S} = \{(\mathbf{I}_i, \mathbf{L}_j)\}_i$ composed of CT images $\mathbf{I}_i$ of the vertebrae and their corresponding labels $\mathbf{L}_i$, with $i \in |T^\mathrm{S}|$. 
We train a segmenter $\mathrm{S}(\mathbf{I}; \theta_\mathrm{S})$ with $\theta_\mathrm{S}$ being the learned parameters for network $\mathrm{S}$.
We use this to generate a per-voxel probability map $\mathbf{P}$ that will associate with each voxel of an image $\mathbf{I}$ the probability of that voxel being part of a vertebra level. 

\begin{figure}[h!]
    \begin{center}
    \begin{tabular}{c}
        \includegraphics[width=\linewidth]{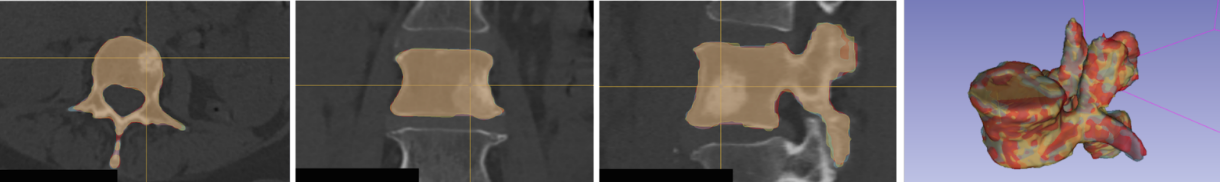}
    \end{tabular}
    \end{center}
    \caption{Our two-step registration pipeline first classifies and segment the vertebra to generate 3D surfaces using, then use a surface-based registration approach to  align the longitudinal CTs.}
    \label{fig:overview}
\end{figure}

The network $\mathrm{S}$ is a deep CNN that follows a standard 3D U-Net architecture \cite{ronneberger2015u} with a $\textsf{Softmax}$ final layer.
Because we are interested in a joint segmentation and classification where we can outline the shape of the vertebra and identify its level, we optimize a \textit{categorical cross-entropy} loss function using gradient descent over the parameters $\theta_\mathcal{S}$ following:
\begin{equation}
L(\theta_\mathrm{S}, T^\mathrm{S}) = \sum_{i \in |T^\mathrm{S}|} \big \Vert \mathrm{S}(\mathbf{I}_i; \theta_\mathrm{S} ) - \mathbf{L}_i \big \Vert_2
\end{equation}
We used the nnU-Net \cite{isensee2021nnu} implementation, a self-adapting framework for 3D full-resolution image segmentation based on the U-Net architecture, to train the network (see Fig. \ref{fig:overview}).

\subsection{Surface-based Longitudinal Registration}
In order to perform the longitudinal CT registration, we used triangulated surfaces reconstructed from the segmentation masks resulting from $\mathrm{S}$.
These surfaces are reconstructed using a marching cubes algorithm, followed by an edge collapse-based incremental decimation \cite{fedorov2015open}. 
We generate a surface for the baseline CT and the 3-months, 6-month and 12-month follow-ups CTs. 
We perform a pair-wise registration, and align the two surfaces by solving a probability density estimation problem \cite{fedorov2015open}. 
The baseline surface represents Gaussian mixture model (GMM) centroids, and each follow-up surface represents data points.
This can be solved by minimizing the negative log-likelihood function based on the GMM approach where we drive the surface of the model using implicit surface-to-surface forces, from baseline to follow-ups. 
Formally, assuming the $N$ observations $\mathbf{x}$ and $M$ GMM centroids $\mathbf{y}$, we aim at finding the transformation of the GMM centroid locations as $\mathcal{T}(\mathbf{y}_m; \mathbf{R}, \mathbf{t}, s) = s\mathbf{R}\mathbf{y}_m + \mathbf{t}$, where $\mathbf{R}$ is a rotation matrix, $\mathbf{t}$ is a translation vector and $s$ is a scaling parameter. 
The objective function to minimize is as follows:
\begin{equation}
E(\mathbf{R}, \mathbf{t}, s, \sigma^2) = \frac{1}{2\sigma^2} \sum^{M,N}_{m,n=1}P(\mathbf{y}_m|\mathbf{x}_n) \Vert \mathbf{x}_n - s\mathbf{R}\mathbf{y}_m - \mathbf{t} \Vert^2 + \frac{3N_P}{2}log(\sigma^2)
\end{equation}
where $P(\cdot)$ denotes the GMM probability density function that smoothly weight the correspondences between the two surfaces.
Finally, this function is optimized using an expectation maximization (EM) algorithm (see Fig. \ref{fig:overview}).

\section{RESULTS}
The segmentation framework was trained on 55 spines using a 5-fold cross-validation strategy. The CTs were acquired from 2 manufacturers in 2 hospitals with an image resolution of 0.35 x 0.35 mm - 1.0 x 1.0 mm in-plane and 0.5 mm - 1.5 mm thickness. We defined 18 classes across vertebral levels from C7 to L5, achieving a validation accuracy of 0.886 (± 0.382). Testing the model on 5 unseen spines yielded an average accuracy of 0.943 (± 0.239), indicating its robust generalization. Then, we used the trained segmentation model to predict the baseline and follow-up 3D surfaces for 5 patients without ground true segmentations. 

\begin{figure}[ht!]
\centering 
\includegraphics[width=\linewidth]{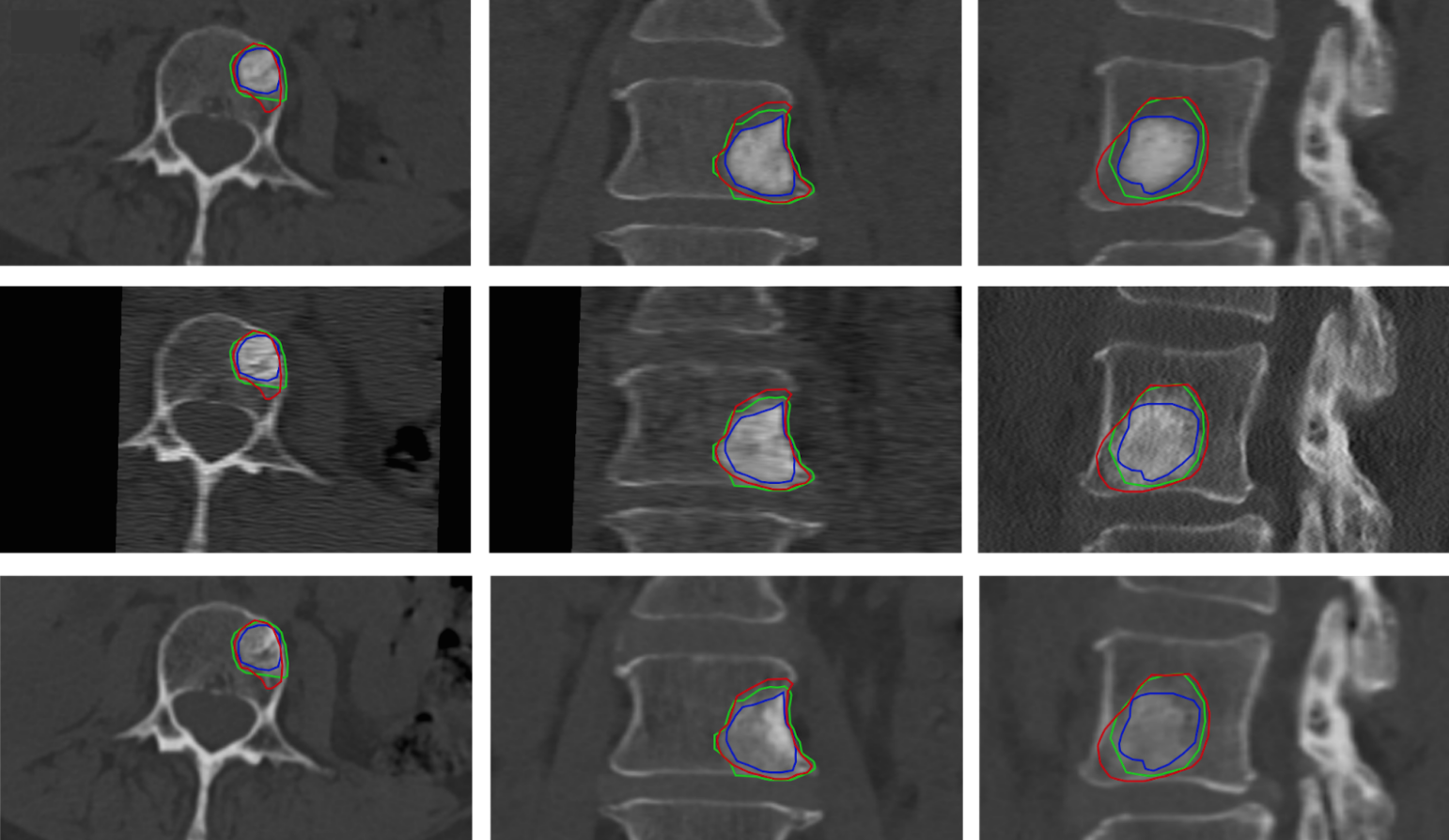}
\caption{Axial, coronal and sagittal planes showing the lesion progression in one case. The lesion is outlined over time (baseline in blue, 3 months in green, 6 months in red) and fused with the longitudinal CT set. Top to bottom: baseline, 3 and 6 months.}
\label{fig:res}
\end{figure}

By registering the 37 predicted vertebrae individually, we performed 111 registrations, aligning 3, 6 and 12 months follow-ups with the baseline images. The mean Dice Similarity Coefficient achieved by our registration method is 0.92±0.11. The Dice score provides a measure of the overlap between the baseline and a follow-up, indicating the accuracy of the registration process. The high mean Dice of 0.92±0.11 signifies a precise alignment between the registered vertebrae. The average Hausdorff distance obtained from our method was measured at 0.65 mm (95\% within 1.74 mm). The Hausdorff distance reflects the discrepancy between the points of two sets, indicating the extent of spatial dissimilarity between the registered vertebrae. Our method's low average Hausdorff distance suggests excellent alignment of the follow-ups into the baseline. The registration is computationally fast with an execution time below 5 seconds. Figure \ref{fig:res} represents our method's results, demonstrating an example of lesion growth over time. Here, an accurate segmentation mask and precise follow-up registration enabled longitudinal lesion tracking in a metastatic vertebrae, including size, shape and structure monitoring.

Importantly, there is a correlation  between segmentation accuracy and subsequent registration performance. As evidenced by our experiments, improved segmentation quality directly influenced the registration Dice score and Hausdorff distance. When segmentation accurately identifies the anatomical boundaries, it provides a more precise representation of the target structures, resulting in improved alignment during the registration process.

\captionsetup[subfigure]{labelformat=empty}

\section{CONCLUSION}
In this paper we presented an automated deep-learning-based tool for registration of longitudinal spine CT. Analysis on the aligned CTs can be used to visually assess and quantify lesion growth and response to treatment. Further studies with larger datasets and different types of spinal diseases are warranted to validate the performance of these algorithms and their potential in predicting disease progression for improved treatment and management.
Future work will first consist of providing shape analysis of the longitudinal lesion growth. 
We will also integrate the type of lesion in the classification model to further monitor and evaluate the lesion growth. 
To extend the evaluation of this approach we are preparing a larger dataset of longitudinal spine CTs. 

\section{ACKNOWLEDGEMENT}
Research reported in this paper was supported by the National Institute of Arthritis and Musculoskeletal and Skin Diseases of the National Institutes of Health under award number R01AR075964.

\bibliography{main}   
\bibliographystyle{spiebib}   

\end{document}